# New Polymorphs of Two-Dimensional Indium Selenide with Enhanced Electronic Properties

*Yuanhui Sun,[†] Yawen Li,[†] Tianshu Li,[†] Koushik Biswas,[‡] Amalia Patanè,[§]*

*and Lijun Zhang[\*,†]*

[†]State Key Laboratory of Integrated Optoelectronics, Key Laboratory of Automobile Materials of MOE and College of Materials Science and Engineering, Jilin University, Changchun 130012, China

[‡]Department of Chemistry and Physics, Arkansas State University, AR 72467, United States

[§]School of Physics and Astronomy, The University of Nottingham, Nottingham NG7 2RD, United Kingdom

*Address correspondence to: lijun_zhang@jlu.edu.cn

# Abstract

The two-dimensional (2D) semiconductor indium selenide (InSe) has attracted significant interest due its unique electronic band structure, high electron mobility and wide tunability of its band gap energy achieved by varying the layer thickness. All these features make 2D InSe a potential candidate for advanced electronic and optoelectronic applications. Here, we report on the discovery of new polymorphs of InSe with enhanced electronic properties. Using a global structure search that combines artificial swarm intelligence with first-principles energetic calculations, we identify polymorphs that consist of a centrosymmetric monolayer belonging to the point group $D_{3d}$, distinct from the well-known polymorphs based on the $D_{3h}$ monolayers that lack inversion symmetry. The new polymorphs are thermodynamically and kinetically stable, and exhibit a wider optical spectral response and larger electron mobilities compared to the known polymorphs. We discuss opportunities to synthesize these newly discovered polymorphs and viable routes to identify them by X-ray diffraction, Raman spectroscopy and second harmonic generation experiments.

# 1. Introduction and background

Atomically thin two-dimensional (2D) layered van der Waals (vdW) semiconductors with high carrier mobility and tunable band gap energy hold promise for next-generation nanoscale electronics and optoelectronics. In particular, indium selenide (InSe) has attracted significant research interest.[1–7] Amongst the 2D layered semiconductors, it exhibits the highest room-temperature electron mobility (~$10^3$ $cm^2V^{-1}s^{-1}$).[1,2,8] This arises from the relatively small effective mass of the conduction band (CB) electrons and dispersive In-$s$ antibonding CB states.[9] Also, it has a direct band gap of ~ 1.25 eV in the bulk form,[10] which increases by nearly 1 eV in monolayer InSe (2.1 eV)[11] due to the cooperative effects of quantum confinement and interlayer coupling.[6,12] Despite its weak optical emission in thin layers[11,12] and the low hole mobility,[3] InSe has already found promising applications in field-effect transistors,[2,3] photodetectors,[5,13,14] and image sensors.[15]

Through mechanical exfoliation[5,12] and epitaxial growth[16] approaches, high-quality InSe layers can be fabricated. To date, three bulk polymorphs (*i.e.*, β, γ, and ε phases) were reported,[2,3,14,17–19] which represent different stacking patterns of the common $D_{3h}$ monolayer structure. In the monolayer In- and Se-atoms are arranged in the sequence of [Se-In-In-Se], forming a graphene-like honeycomb lattice but without inversion symmetry, in which vertically aligned In-Se atomic pairs occupy different sublattice sites. It is known that many 2D materials possess different structures of the monolayer, such as graphene and graphdiyne allotropes of monolayer carbon,[20,21] trigonal prismatic (2H) and octahedral (1T) monolayers of transition metal dichacolcogenides,[22,23] black phosphorene and blue phosphorene of monolayer phosphorus.[24,25] Different monolayer structures endow 2D materials with variable properties useful for different applications. For example, the 2H-monolayer derived phases of $MoS_2$ are promising semiconductors for electronic and optoelectronic devices,[26–30] whereas the 1T-monolayer derived phases are conductive metals for electrocatalytic[31,32] and energy storage[23] applications. The polymorphic nature of the monolayers is clearly an enticing feature of many 2D materials. However, to date, only the $D_{3h}$ monolayer structure was reported for InSe.[1–19] Thus, it may be worthwhile to explore the existence of other InSe monolayers with potentially emergent properties.

Here, we report on a new monolayer polymorph of InSe. Through a global structure search study that combines artificial swarm intelligence with first-principles energetic calculations, we identify a new monolayer polymorph that belongs to the point group $D_{3d}$ with inversion symmetry, distinct from the known non-centrosymmetric $D_{3h}$ monolayer. It is thermodynamically comparable in energy with the $D_{3h}$, showing robust phonon and thermal stability, as well as kinetic stability with respect to a transformation to $D_{3h}$. Three bulk phases based on different stacks of the $D_{3d}$ monolayer are predicted, one of which show an enhanced band gap tunability with varying layer thickness and higher electron mobility compared to the other phases. We discuss how the new phases could be identified by X-ray diffraction (XRD), Raman spectroscopy and second harmonic generation (SHG) measurements, thus opening realistic prospects for the experimental observation and investigation of the predicted polymorphs.

## 2. Results and discussion

**Searching for stable polymorphs of InSe with swarm intelligence guided structure searches.** We performed global structure searches at the chemical composition of In:Se=1:1, which involves nearly 3600 structure points sampled from the free energy landscape. Figure 1a shows the evolution of the energy of the sampled structures as a function of search generation. The zoomed-in low-energy region is dominated by three types of polymorphs (green, red, and blue dots in Figure 1a), each of which is composed of the common monolayer structure (shown in corresponding colored box of Figure 1b). The higher-energy structures consist of the monolayer in a modified version of the InSe structure (point group $C_{2h}$, Figure 1b, upper panel), a high-pressure phase discovered experimentally.[33] It is actually a non-layered phase with strong covalent bonding. The two low-energy types (blue and red dots in Figure 1a) have comparable energy within several meV/atom. The experimentally known non-centrosymmetric $D_{3h}$ monolayer constitutes the polymorphs shown in Figure 1b middle panel, corresponding to the β($D_{3h}$),[34] γ($D_{3h}$),[35] or ε($D_{3h}$)[36] phases with symmetries of $P6_3/mmc$, $R3m$, and $P\bar{6}m2$, respectively. The polymorphs shown in the bottom panel of Figure 1b are composed of a strikingly new monolayer with point group $D_{3d}$. Its graphene-like honeycomb lattice is occupied by In-Se atomic pairs along the vertical direction with inversion symmetry, distinct from the vertically aligned In-Se atomic pairs with mirror symmetry and point group $D_{3h}$. The centrosymmetric feature of the $D_{3d}$ monolayer endows its constituted polymorphs with emergent electronic and optical properties, as described below.

By systematically analyzing the structure search results and additional calculations of candidate stacks based on the $D_{3d}$ monolayer, we found three energetically favorable bulk phases, named as δ($D_{3d}$), ω($D_{3d}$), and φ($D_{3d}$) (Figure 1c, with explicit structural information given in Table S1 of the Supporting Information). They are in space groups $P\bar{3}m1$, $P6_3mc$, and $R\bar{3}m$, featuring vertical stacking patterns AA, AB (with adjacent monolayers in-plane rotated by 180º and shifted by 1/3 unit), and ABC (with monolayers shifted by 1/3 unit), respectively. Comprehensive calculations using different vdW functionals indicate that they have comparable energies within 5 meV/atom (Figure S1 of the Supporting Information), of which the φ($D_{3d}$) phase is energetically most favored. Comparing the new δ($D_{3d}$), ω($D_{3d}$), and φ($D_{3d}$) phases with the experimentally known ones of the $D_{3h}$ monolayer [β($D_{3h}$), γ($D_{3h}$), and ε($D_{3h}$)], we found that the explicit energy differences are less than 8 meV/atom regardless of the specific vdW functional used. The energy difference is much smaller than that between the 2H and 1T phases of $MoS_2$ (~70 meV/atom),[37] or of that between black and blue phosphorene (~14 meV/atom).[38] This thermodynamically favorable condition implies that there is reasonable likelihood of synthesizing the newly found InSe polymorphs.

**Robust kinetic, thermal and phonon stability of the newly found polymorphs.** Even though one material phase is thermodynamically favored, its kinetic transformation into another favored phase may prohibit its stabilization. Kinetic stability of the new $D_{3d}$ InSe monolayer is examined by calculating the transition barrier between $D_{3d}$ and the known $D_{3h}$ monolayer. By identifying feasible transition pathways involving a saddle-point transition state, we obtained a quite large activation barrier of ~150 meV/atom (Figure 2a). This indicates that the $D_{3d}$ monolayer can be kinetically stabilized once it is synthesized. We further examined the thermal stability of the new $D_{3d}$ monolayer using molecular dynamics simulations at 300 K and 500 K (Figure 2b). All atoms vibrate around their equilibrium positions at both temperatures and the In-Se network of the monolayer is retained. The time-dependent energy fluctuation at 500 K is expectedly larger, but the equilibrium structure remains anchored to the $D_{3d}$ symmetry. Good thermal stability above room temperature indicates the possibility of growing $D_{3d}$ monolayer based polymorphs via controlled high-temperature methods, such as chemical vapor deposition that has been recently used to grow ultrathin InSe flakes.[39] Finally, we examined the lattice dynamical stability by calculating the phonon spectrum of the $D_{3d}$ monolayer. As shown in Figure 2c, the dynamical stability of the $D_{3d}$ monolayer lattice is evidenced by the absence of imaginary phonon modes at 0K. The phonon spectra of the δ($D_{3d}$), ω($D_{3d}$), and φ($D_{3d}$) bulk polymorphs were also calculated (Figure 2d). All polymorphs demonstrate robust phonon stability. Although the phonon spectra of the three polymorphs look similar, the shear (in red) and breathing (in blue) mode branches show substantial differences, indicating different degrees of interlayer coupling, as discussed further below.

**Emergent electronic properties of the $D_{3d}$ monolayer based polymorphs.** As shown in Figure 3a, the electronic band structure of the new $D_{3d}$ monolayer show similar band-edge states in proximity to the Γ point as for the known $D_{3h}$ monolayer. The dispersive conduction band minimum is located at Γ. Two nearly degenerate weakly dispersed valence band maxima (VBM) are observed along the Γ–M and Γ–K directions. Thus, both $D_{3h}$ and $D_{3d}$ monolayers are indirect band gap semiconductors. Both conduction and valence band-edge states are dominated by the Se-$p_z$ orbital. The band gap energy calculated with the hybrid functional approach[40] is 2.30 eV, slightly smaller than the value of 2.39 eV for the $D_{3h}$ monolayer. By stacking the $D_{3d}$ monolayers into bulk $\delta(D_{3d})$, $\omega(D_{3d})$, and $\varphi(D_{3d})$ polymorphs, the band gap changes from indirect to direct and its value decreases (Figure S2-S4 of the Supporting Information). Figure 3b shows the evolution of the band gap energy with increasing number of layers for the three polymorphs and for the $D_{3h}$ monolayer based $\beta(D_{3h})$ and $\gamma(D_{3h})$ polymorphs. To remedy the band gap underestimation issue of the density functional theory (DFT) calculations, we adopted a scissor operator with magnitude equal to the band gap energy difference between the DFT and hybrid functional calculations for the $D_{3d}/D_{3h}$ monolayer to all the multiple-layered cases. The resulting band gaps for the $\delta(D_{3d})$, $\omega(D_{3d})$, and $\varphi(D_{3d})$ phases are 0.81 eV, 0.96 eV, and 1.05 eV, respectively. Amongst all phases, bulk $\delta(D_{3d})$ shows the smallest gap value, which is beneficial for applications that require an optical response in the infrared spectral range. Also, with varying the layer thickness the $\delta(D_{3d})$ phase spans a wider energy range of energy gaps (~1.5 eV) than the known β($D_{3h}$) and γ($D_{3h}$) polymorphs (~1.2-1.3 eV). The decrease in the band gap from monolayer to multiple layers InSe is predominantly caused by the upward shift of the VBM (Figure S5 of the Supporting Information). Amongst all polymorphs, the δ($D_{3d}$) phase shows the most dramatic change in the VBM energy, consistent with its largest band gap variation (Figure 3b). This originates from its shortest interlayer distance (8.27 Å, Figure 3c) associated with the $D_{3d}$ symmetry and the specific layer stacking pattern. As a result, the stronger interlayer coupling occurs in the δ($D_{3d}$) phase, which is evidenced by the stronger directional charge overlapping/chemical bonding between the Se-atoms in adjacent layers (Figure 3c). This is responsible for the more pronounced upward shift of the VBM and thus the larger band gap change.

The calculated electron mobility of the new $D_{3d}$ monolayer is 912 $cm^2V^{-1}s^{-1}$ and 945 $cm^2V^{-1}s^{-1}$ along zigzag and armchair directions, respectively, higher than the values of 689 $cm^2V^{-1}s^{-1}$ and 801 $cm^2V^{-1}s^{-1}$ of the $D_{3h}$ monolayer calculated in the same approach (Table S2 of the Supporting Information). The increase in mobility originates primarily from the smaller electron effective masses (0.18 $m_0$ and 0.19 $m_0$) of the $D_{3d}$ monolayer compared with those of the $D_{3h}$ monolayer (0.20 $m_0$ and 0.23 $m_0$). With increasing number of layers, the δ($D_{3d}$), ω($D_{3d}$), and φ($D_{3d}$) phases show remarkably enhanced electron mobility, reaching values of 9500-13000 $cm^2V^{-1}s^{-1}$ and 10000-14000 $cm^2V^{-1}s^{-1}$ for 6 monolayers along zigzag and armchair directions, respectively (Figure 4). This resembles the behavior of the existing β($D_{3h}$) and γ($D_{3h}$) polymorphs,[6,41] which we ascribe to the decreasing electron effective mass (Table S2 of the Supporting Information) and a likely increase in the carrier scattering time due to reduced electron-phonon coupling in the multiple layers. Here, the interlayer coupling or interaction plays an essential role. We note that for a given layer thickness the δ($D_{3d}$), ω($D_{3d}$), and φ($D_{3d}$) phases exhibit a higher electron mobility than the β($D_{3h}$) and γ($D_{3h}$) phases. In particular, the δ($D_{3d}$) phase, which has the strongest interlayer coupling, exhibits the highest electron mobility: for 6 monolayers, the mobility is enhanced by a factor of about 1.4 and 1.6 along zigzag and armchair directions with respect to the β($D_{3h}$) phase. Therefore, compared to polymorphs based on the $D_{3h}$ monolayer, polymorphs based on the $D_{3d}$ monolayer have more favorable properties for electronic and optoelectronic applications.

**Experimental identification of the predicted $D_{3d}$ monolayer based polymorphs.** XRD,[2,12,16] Raman spectroscopy,[13,14,42] and SHG measurements[18,39] have been recently used to identify the $D_{3h}$ monolayer based β($D_{3h}$) and γ($D_{3h}$) polymorphs. We suggest that the same characterization approaches can be adopted to identify the predicted $D_{3d}$ monolayer based polymorphs and distinguish them from those based on the known $D_{3h}$ monolayer. Figure 5a shows the simulated XRD patterns of δ($D_{3d}$), ω($D_{3d}$), and φ($D_{3d}$) polymorphs, compared with the experimental powder diffraction file (PDF) of β($D_{3h}$) and γ($D_{3h}$). While all polymorphs show two common low-angle peaks at about 10.6º and 21.2º, the δ($D_{3d}$) and ω($D_{3d}$) phases show a rather different peak distribution between 25º and 30º with respect to β($D_{3h}$) and γ($D_{3h}$). For the φ($D_{3d}$) phase, the XRD pattern is quite similar to that of γ($D_{3h}$), with a slight shift towards low-angles. However, as shown in Figure 5b, the φ($D_{3d}$) and γ($D_{3h}$) phases can be distinguished by their Raman-active phonon modes. The two phases have similar Raman phonons in the low- and high-frequency ranges; in contrast, between 150 cm$^{-1}$ and 200 cm$^{-1}$, they show distinct features: for γ($D_{3h}$), there are two nearly degenerate $E$-modes and one $A_1$-mode, but there is only one $E_g$-mode in φ($D_{3d}$). We note that the existence of the $A_1$ mode in γ($D_{3h}$) has not been always reported in InSe,[12–14,42,43] which in retrospect may be a hint towards the identification of the φ($D_{3d}$) phase. In addition, since the new $D_{3d}$ monolayer has inversion symmetry, no SHG response should be observed, which is distinct from the SHG response of the non-centrosymmetric $D_{3h}$ monolayer.[44]

By stacking the $D_{3d}$ monolayers into the $\delta(D_{3d})$, $\omega(D_{3d})$, and $\varphi(D_{3d})$ polymorphs, a SHG response may emerge. This would depend on the specific stacking pattern and number of layers (Table S3 of the Supporting Information), providing a means of identifying the predicted $D_{3d}$ monolayer based polymorphs.

## 3. Conclusion

To conclude, we explored via swarm-intelligence based computational structure searches the free energy landscape of 2D layered semiconductor InSe investigate the possibility of new polymorphs. In addition to the existing ambient and high-pressure polymorphs, three yet to be reported polymorphs were identified, which consist of a new centrosymmetric monolayer with point group $D_{3d}$, distinct from the polymorphs built from noncentrosymmetric $D_{3h}$ monolayer. The new $D_{3d}$ monolayer based polymorphs show thermodynamic stability, comparable to that of known $D_{3h}$ monolayer based ones, as indicated by the small energy difference (< 10 meV/atom). The new polymorphs are kinetically stable against transformation to the known $D_{3h}$ monolayer based polymorphs, demonstrate dynamical stability, and robust thermal stability at room and high temperatures. The $D_{3d}$ monolayer based polymorphs have indirect gaps with band gap energies that vary widely from the monolayer to bulk by up to ~1.5 eV, higher than in the $D_{3h}$ monolayer based polymorphs calculated in the same way. In addition, for the same layer thickness the $D_{3d}$ monolayer based polymorphs exhibit higher electron mobility, with an enhancement factor of up to ~1.6. These properties arise from the stronger interlayer electronic coupling associated with the $D_{3d}$ symmetry and the specific layer stacking pattern. Finally, the predicted new polymorphs can be distinguished from the known phases by X-ray diffraction, Raman spectroscopy and second harmonic generation measurements. Our prediction of new $D_{3d}$ monolayer based InSe polymorphs with enhanced electronic properties offers prospects for further research on the synthesis and exploitation of new promising 2D materials with different polymorphic nature.

## 4. Computational approaches

The search for the polymorphs of 2D InSe was carried out using a global minimum search of the free energy landscape with respect to structural variations by combining a particle swarm optimization algorithm with first-principles energetic calculations.[45,46] With this methodology one can find the ground-state or metastable structures based on the known chemical composition, without relying on any prior known structural information. Its validity in crystal structure search has been demonstrated in a variety of material systems,[47–52] including 2D layered materials.[53–55] The underlying first-principles DFT calculations were carried out by using the plane-wave pseudopotential method as implemented in Vienna *ab initio* Simulation Package.[56,57] The electron-ion interactions were described by the projected augmented wave (PAW) pseudopotentials[58,59] with In-$5s^25p^1$ and Se-$4s^24p^4$ treated as valence electrons. We used the generalized gradient approximation formulated by Perdew, Burke, and Ernzerhof[60] as exchange-correlation functional. We adopted a kinetic energy cutoff of 520 eV for wave-function expansion and a $k$-point mesh of $2\pi \times 0.03$ Å$^{-1}$ or less for Brillouin zone integration. A vacuum layer of more than 15 Å thickness was used in layer-dependent calculations to isolate the InSe layer from its neighboring image. The structures (lattice parameters and atomic positions) were fully optimized including vdW interaction, until the residual forces were converged within 0.02 eV/Å. The optB86b-vdW functional[61,62] was adopted, that previously provided a good description of the structural properties of the known β, γ InSe phases.[6] We employed the hybrid functional approach[40] (with 25% exact Fock exchange) to remedy the band gap underestimation in DFT based calculations. The transition barriers between the monolayer polymorphs were calculated using the nudged elastic band method in conjunction with the climbing image method.[63,64] *Ab initio* molecular dynamics simulations were performed at 300 K and 500 K using the NVT ensemble and the temperature was controlled by using the Nosé-Hoover method.[65] Phonon spectra were calculated using a finite-difference supercell approach[66] implemented in the Phonopy code.[67] The layer-dependent carrier mobility was evaluated within the deformation potential theory,[68,69] using carrier effective mass along transport direction and in-plane elastic modulus as inputs. A more detailed description of the computational procedures can be found in the Supplementary Information.

**Acknowledgements**

This work is supported by the National Natural Science Foundation of China (Grants No. 61722403 and 11674121) and Jilin Province Science and Technology Development Program (Grant No. 20190201016JC). Calculations were performed in part at the high performance computing center of Jilin University. We are grateful to Prof. Y. Ye for helpful discussion. AP acknowledges the Chinese Academy of Sciences (CAS) for the Award of a "President's International Fellowship for Visiting Scientists" and the European Union's Horizon 2020 research and innovation programme Graphene Flagship Core 3.

**Keywords**

2D materials, Indium Selenide, electronics and optoelectronics, materials by design

# References

[1] D. A. Bandurin, A. V. Tyurnina, G. L. Yu, A. Mishchenko, V. Zolyomi, S. V. Morozov, R. K. Kumar, R. V. Gorbachev, Z. R. Kudrynskyi, S. Pezzini, Z. D. Kovalyuk, U. Zeitler, K. S. Novoselov, A. Patane, L. Eaves, I. V. Grigorieva, V. I. Fal'ko, A. K. Geim, Y. Cao, *Nat. Nanotechnol.* **2017**, *12*, 223.
[2] W. Feng, W. Zheng, W. Cao, P. Hu, *Adv. Mater.* **2014**, *26*, 6587.
[3] S. Sucharitakul, N. J. Goble, U. R. Kumar, R. Sankar, Z. A. Bogorad, F. C. Chou, Y. T. Chen, X. P. Gao, *Nano Lett.* **2015**, *15*, 3815.
[4] Mudd Garry W., Svatek Simon A., Hague Lee, Makarovsky Oleg, Kudrynskyi Zakhar R., Mellor Christopher J., Beton Peter H., Eaves Laurence, Novoselov Kostya S., Kovalyuk Zakhar D., Vdovin Evgeny E., Marsden Alex J., Wilson Neil R., Patanè Amalia, *Adv. Mater.* **2015**, *27*, 3760.
[5] S. R. Tamalampudi, Y.-Y. Lu, R. Kumar U., R. Sankar, C.-D. Liao, K. Moorthy B., C.-H. Cheng, F. C. Chou, Y.-T. Chen, *Nano Lett.* **2014**, *14*, 2800.
[6] Y. Sun, S. Luo, X.-G. Zhao, K. Biswas, S.-L. Li, L. Zhang, *Nanoscale* **2018**, *10*, 7991.
[7] Q. Peng, R. Xiong, B. Sa, J. Zhou, C. Wen, B. Wu, M. Anpo, Z. Sun, *Catal. Sci. Technol.* **2017**, *7*, 2744.
[8] A. Segura, F. Pomer, A. Cantarero, W. Krause, A. Chevy, *Phys. Rev. B* **1984**, *29*, 5708.
[9] P. Gomes da Costa, R. G. Dandrea, R. F. Wallis, M. Balkanski, *Phys. Rev. B* **1993**, *48*, 14135.
[10] J. Camassel, P. Merle, H. Mathieu, A. Chevy, *Phys. Rev. B* **1978**, *17*, 4718.
[11] M. Brotons-Gisbert, D. Andres-Penares, J. Suh, F. Hidalgo, R. Abargues, P. J. Rodriguez-Canto, A. Segura, A. Cros, G. Tobias, E. Canadell, P. Ordejon, J. Wu, J. P. Martinez-Pastor, J. F. Sanchez-Royo, *Nano Lett.* **2016**, *16*, 3221.
[12] G. W. Mudd, S. A. Svatek, T. Ren, A. Patane, O. Makarovsky, L. Eaves, P. H. Beton, Z. D. Kovalyuk, G. V. Lashkarev, Z. R. Kudrynskyi, A. I. Dmitriev, *Adv. Mater.* **2013**, *25*, 5714.
[13] Z. Chen, J. Biscaras, A. Shukla, *Nanoscale* **2015**, *7*, 5981.
[14] S. Lei, L. Ge, S. Najmaei, A. George, R. Kappera, J. Lou, M. Chhowalla, H. Yamaguchi, G. Gupta, R. Vajtai, A. D. Mohite, P. M. Ajayan, *ACS Nano* **2014**, *8*, 1263.
[15] S. Lei, F. Wen, B. Li, Q. Wang, Y. Huang, Y. Gong, Y. He, P. Dong, J. Bellah, A. George, L. Ge, J. Lou, N. J. Halas, R. Vajtai, P. M. Ajayan, *Nano Lett.* **2015**, *15*, 259.
[16] Z. Yang, W. Jie, C.-H. Mak, S. Lin, H. Lin, X. Yang, F. Yan, S. P. Lau, J. Hao, *ACS Nano* **2017**, *11*, 4225.
[17] Han Guang, Chen Zhi‐Gang, Drennan John, Zou Jin, *Small* **2014**, *10*, 2747.
[18] Q. Hao, H. Yi, H. Su, B. Wei, Z. Wang, Z. Lao, Y. Chai, Z. Wang, C. Jin, J. Dai, W. Zhang, *Nano Lett.* **2019**, *19*, 2634.
[19] D. W. Boukhvalov, B. Gürbulak, S. Duman, L. Wang, A. Politano, L. S. Caputi, G. Chiarello, A. Cupolillo, *Nanomaterials* **2017**, *7*, 372.
[20] H. P. Boehm, R. Setton, E. Stumpp, *Carbon* **1986**, *24*, 241.
[21] R. H. Baughman, H. Eckhardt, M. Kertesz, *J. Chem. Phys.* **1987**, *87*, 6687.
[22] Y.-H. Lee, X.-Q. Zhang, W. Zhang, M.-T. Chang, C.-T. Lin, K.-D. Chang, Y.-C. Yu, J. T.-W. Wang, C.-S. Chang, L.-J. Li, T.-W. Lin, *Adv. Mater.* **2012**, *24*, 2320.
[23] M. Acerce, D. Voiry, M. Chhowalla, *Nat. Nanotechnol.* **2015**, *10*, 313.
[24] L. Li, Y. Yu, G. J. Ye, Q. Ge, X. Ou, H. Wu, D. Feng, X. H. Chen, Y. Zhang, *Nat. Nanotechnol.* **2014**, *9*, 372.
[25] J. L. Zhang, S. Zhao, C. Han, Z. Wang, S. Zhong, S. Sun, R. Guo, X. Zhou, C. D. Gu, K. D. Yuan, Z. Li, W. Chen, *Nano Lett.* **2016**, *16*, 4903.

[26] RadisavljevicB., RadenovicA., BrivioJ., GiacomettiV., KisA., *Nat Nano* **2011**, *6*, 147.
[27] J. Chang, S. Larentis, E. Tutuc, L. F. Register, S. K. Banerjee, *Appl. Phys. Lett.* **2014**, *104*, 141603.
[28] Q. H. Wang, K. Kalantar-Zadeh, A. Kis, J. N. Coleman, M. S. Strano, *Nat Nano* **2012**, *7*, 699.
[29] M. Chhowalla, H. S. Shin, G. Eda, L.-J. Li, K. P. Loh, H. Zhang, *Nat Chem* **2013**, *5*, 263.
[30] G. Eda, S. A. Maier, *ACS Nano* **2013**, *7*, 5660.
[31] Y. Huang, Y. Sun, X. Zheng, T. Aoki, B. Pattengale, J. Huang, X. He, W. Bian, S. Younan, N. Williams, J. Hu, J. Ge, N. Pu, X. Yan, X. Pan, L. Zhang, Y. Wei, J. Gu, *Nat. Commun.* **2019**, *10*, 982.
[32] J. Zhang, T. Wang, P. Liu, S. Liu, R. Dong, X. Zhuang, M. Chen, X. Feng, *Energy Environ. Sci.* **2016**, *9*, 2789.
[33] Y. Watanabe, H. Iwasaki, N. Kuroda, Y. Nishina, *J. Solid State Chem.* **1982**, *43*, 140.
[34] B. Čelustka, S. Popović, *J. Phys. Chem. Solids* **1974**, *35*, 287.
[35] J. Rigoult, A. Rimsky, A. Kuhn, *Acta Crystallogr. B* **1980**, *36*, 916.
[36] W. P. Muschinsky, N. M. Pawlenko, *Krist. Tech.* **1969**, *4*, K5.
[37] H. Shu, F. Li, C. Hu, P. Liang, D. Cao, X. Chen, *Nanoscale* **2016**, *8*, 2918.
[38] Z. Zhu, D. Tománek, *Phys. Rev. Lett.* **2014**, *112*, 176802.
[39] W. Huang, L. Gan, H. Li, Y. Ma, T. Zhai, *Chem. – Eur. J.* **2018**, *24*, 15678.
[40] A. V. Krukau, O. A. Vydrov, A. F. Izmaylov, G. E. Scuseria, *J. Chem. Phys.* **2006**, *125*, 224106.
[41] M. Wu, J. Shi, M. Zhang, Y. Ding, H. Wang, Y. Cen, J. Lu, *Nanoscale* **2018**, *10*, 11441.
[42] Z. Chen, K. Gacem, M. Boukhicha, J. Biscaras, A. Shukla, *Nanotechnology* **2013**, *24*, 415708.
[43] N. Kuroda, Y. Nishina, *Solid State Commun.* **1978**, *28*, 439.
[44] J. Zhou, J. Shi, Q. Zeng, Y. Chen, L. Niu, F. Liu, T. Yu, K. Suenaga, X. Liu, J. Lin, Z. Liu, *2D Mater.* **2018**, *5*, 025019.
[45] Y. Wang, J. Lv, L. Zhu, Y. Ma, *Phys. Rev. B* **2010**, *82*, 094116.
[46] Y. Wang, J. Lv, L. Zhu, Y. Ma, *Comput. Phys. Commun.* **2012**, *183*, 2063.
[47] F. Peng, M. Miao, H. Wang, Q. Li, Y. Ma, *J. Am. Chem. Soc.* **2012**, *134*, 18599.
[48] L. Zhu, Z. Wang, Y. Wang, G. Zou, H. Mao, Y. Ma, *Proc. Natl. Acad. Sci.* **2012**, *109*, 751.
[49] Q. Li, D. Zhou, W. Zheng, Y. Ma, C. Chen, *Phys. Rev. Lett.* **2013**, *110*, 136403.
[50] H. Wang, J. S. Tse, K. Tanaka, T. Iitaka, Y. Ma, *Proc. Natl. Acad. Sci.* **2012**, *109*, 6463.
[51] L. Zhu, H. Wang, Y. Wang, J. Lv, Y. Ma, Q. Cui, Y. Ma, G. Zou, *Phys. Rev. Lett.* **2011**, *106*, 145501.
[52] J. Lv, Y. Wang, L. Zhu, Y. Ma, *Phys. Rev. Lett.* **2011**, *106*, 015503.
[53] T. Yu, Z. Zhao, L. Liu, S. Zhang, H. Xu, G. Yang, *J. Am. Chem. Soc.* **2018**, *140*, 5962.
[54] B. Luo, Y. Yao, E. Tian, H. Song, X. Wang, G. Li, K. Xi, B. Li, H. Song, L. Li, *Proc. Natl. Acad. Sci.* **2019**, *116*, 17213.
[55] T. Yu, Z. Zhao, Y. Sun, A. Bergara, J. Lin, S. Zhang, H. Xu, L. Zhang, G. Yang, Y. Liu, *J. Am. Chem. Soc.* **2019**, *141*, 1599.
[56] G. Kresse, J. Furthmüller, *Comput. Mater. Sci.* **1996**, *6*, 15.
[57] G. Kresse, J. Furthmüller, *Phys. Rev. B* **1996**, *54*, 11169.
[58] G. Kresse, D. Joubert, *Phys. Rev. B* **1999**, *59*, 1758.
[59] P. E. Blöchl, *Phys. Rev. B* **1994**, *50*, 17953.
[60] J. P. Perdew, K. Burke, M. Ernzerhof, *Phys. Rev. Lett.* **1996**, *77*, 3865.
[61] J. Klimeš, D. R. Bowler, A. Michaelides, *Phys. Rev. B* **2011**, *83*, 195131.
[62] Jiří Klimeš and David R Bowler and Angelos Michaelides, *J. Phys. Condens. Matter* **2010**, *22*, 022201.
[63] G. Henkelman, B. P. Uberuaga, H. Jónsson, *J. Chem. Phys.* **2000**, *113*, 9901.

[64] G. Henkelman, H. Jónsson, *J. Chem. Phys.* **2000**, *113*, 9978.
[65] G. J. Martyna, M. L. Klein, M. Tuckerman, *J. Chem. Phys.* **1992**, *97*, 2635.
[66] K. Parlinski, Z. Q. Li, Y. Kawazoe, *Phys. Rev. Lett.* **1997**, *78*, 4063.
[67] A. Togo, F. Oba, I. Tanaka, *Phys. Rev. B* **2008**, *78*, 134106.
[68] J. Bardeen, W. Shockley, *Phys. Rev.* **1950**, *80*, 72.
[69] P. J. Price, *Ann. Phys.* **1981**, *133*, 217.

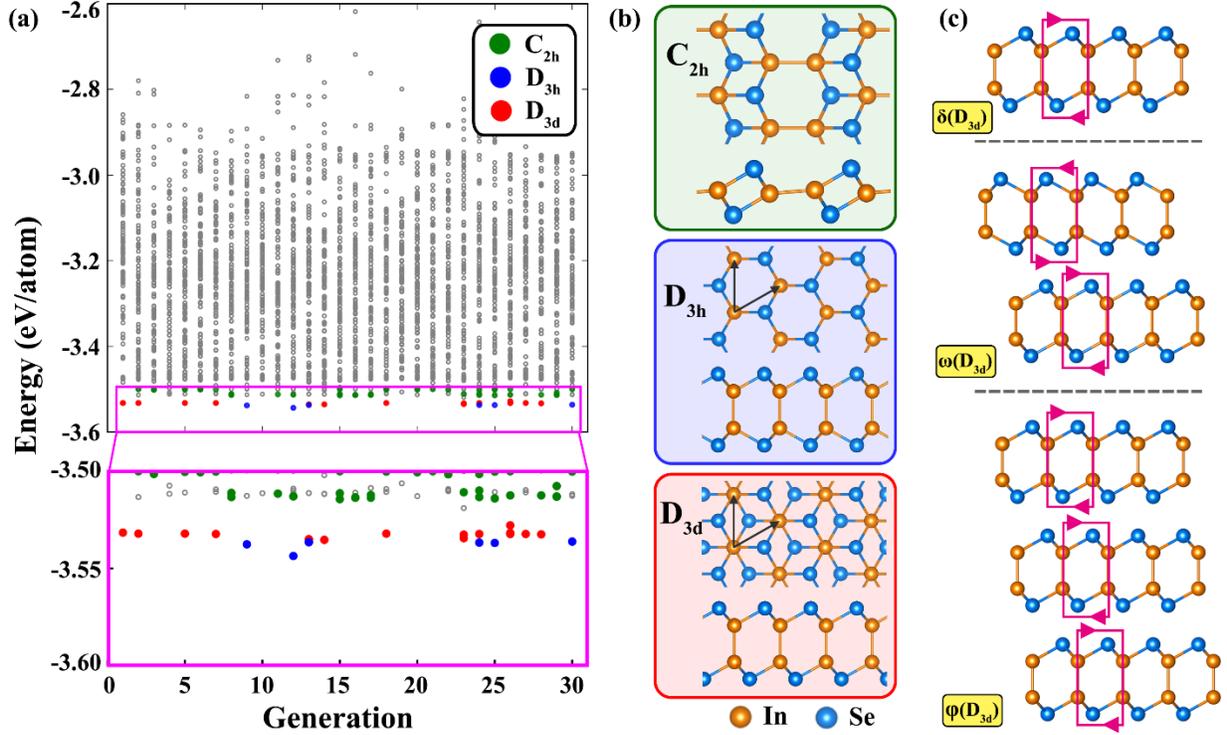

**Figure 1.** (a) Evolution of the energy of the predicted structures as a function search generation. The zoomed-in low-energy region is dominated by three types of polymorphs (shown in green, red, and blue dots). (b) The monolayer structures of three types of polymorphs ($C_{2h}$, $D_{3h}$, and $D_{3d}$) shown in the corresponding colored boxes. (c) Side views of the three energetically favorable bulk phases, named as $\delta(D_{3d})$, $\omega(D_{3d})$, and $\varphi(D_{3d})$, respectively. The rectangular frames with arrows are a visual aid to the particular stacking pattern of the layers.

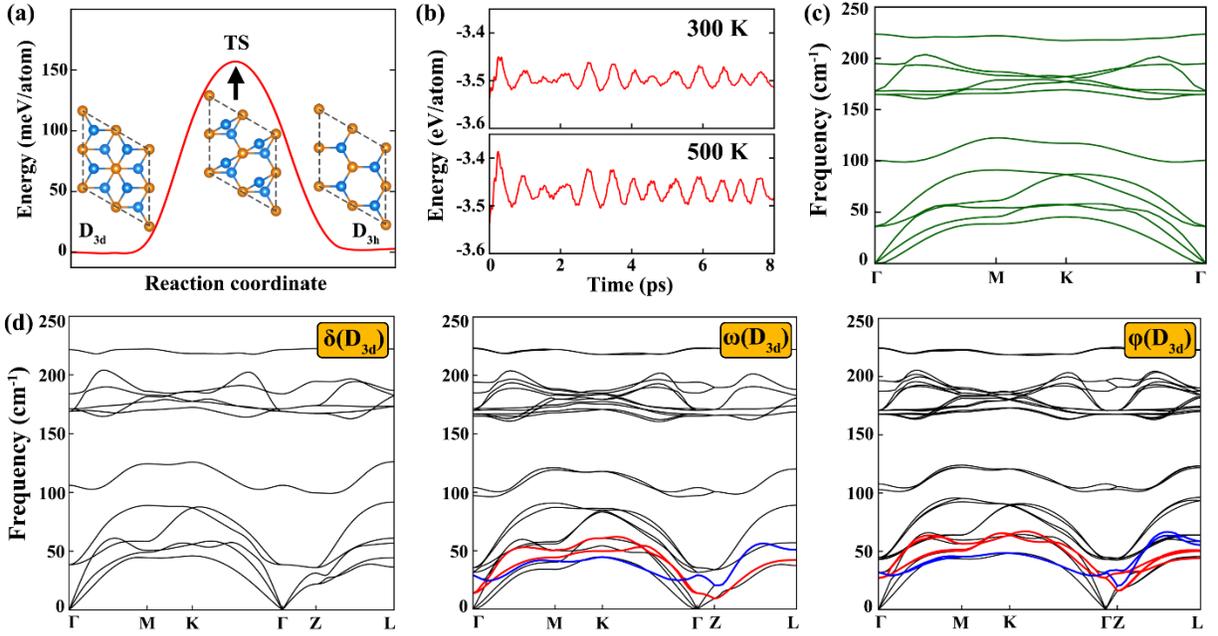

**Figure 2.** (a) Energy barrier and atomic structures during the transformation from the $D_{3d}$ to the $D_{3h}$ monolayer. The transition-state (TS) structure is indicated. (b) Fluctuations of the total potential energy of $D_{3d}$ monolayer during the molecular dynamics simulation at 300 K and 500 K, respectively. (c) Phonon spectrum of the $D_{3d}$ monolayer. (d) Phonon spectra of δ($D_{3d}$), ω($D_{3d}$), and φ($D_{3d}$). The shear and breathing mode branches are shown in red and blue, respectively.

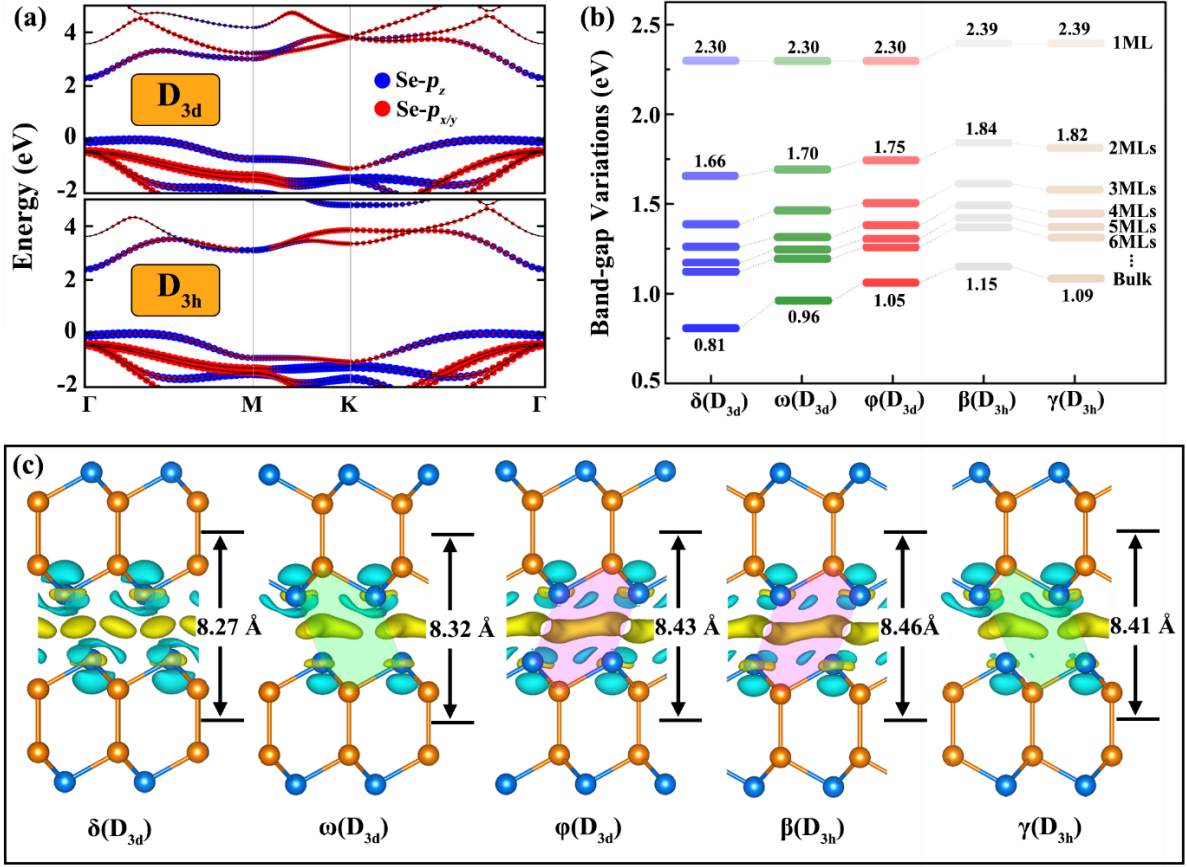

**Figure 3.** (a) Calculated band structures of the $D_{3d}$ and $D_{3h}$ monolayers. The band structures have been projected onto atomic orbitals with the blue color representing Se-$p_z$ and red representing Se-$p_{x/y}$. (b) Evolution of the band gap with increasing number of layers of the predicted three $D_{3d}$ monolayer based polymorphs and two known $D_{3h}$ monolayer based polymorphs. (c) Interlayer differential charge densities of bilayer $\delta(D_{3d})$, $\omega(D_{3d})$, $\varphi(D_{3d})$, $\beta(D_{3h})$, and $\gamma(D_{3h})$, respectively. The isosurface value is set to $1\times10^{-4}$ electrons/Å$^3$. The charge accumulation and depletion are shown in yellow and blue, respectively.

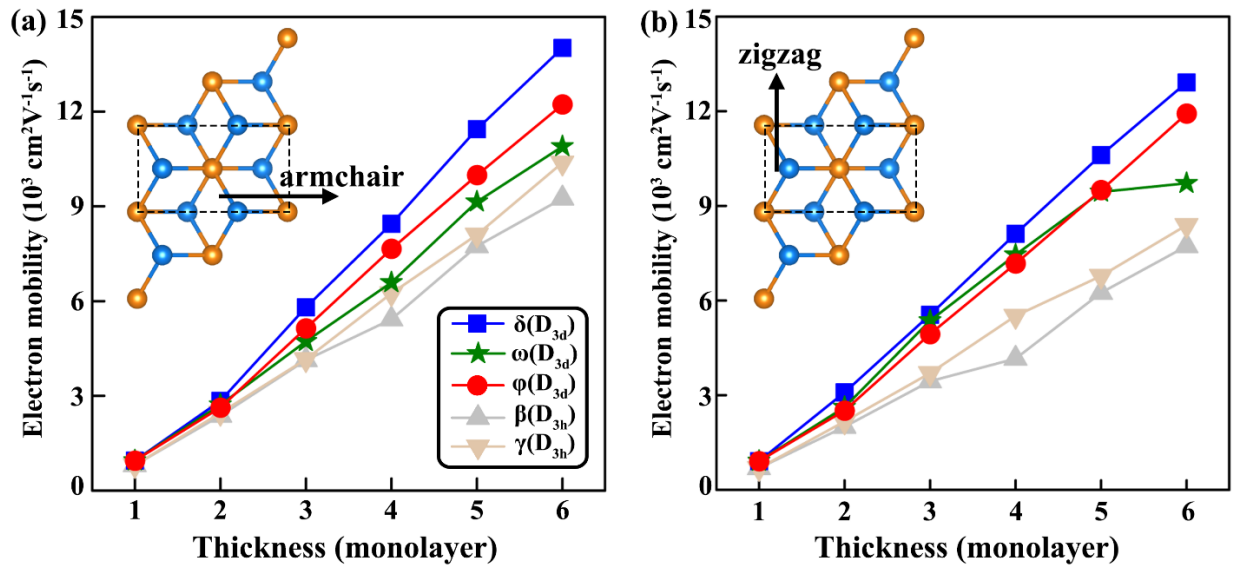

**Figure 4.** Evolution of electron mobilities along the (a) armchair and (b) zigzag directions with increasing number of layers for three $D_{3d}$ monolayer based polymorphs and two known $D_{3h}$ monolayer based polymorphs.

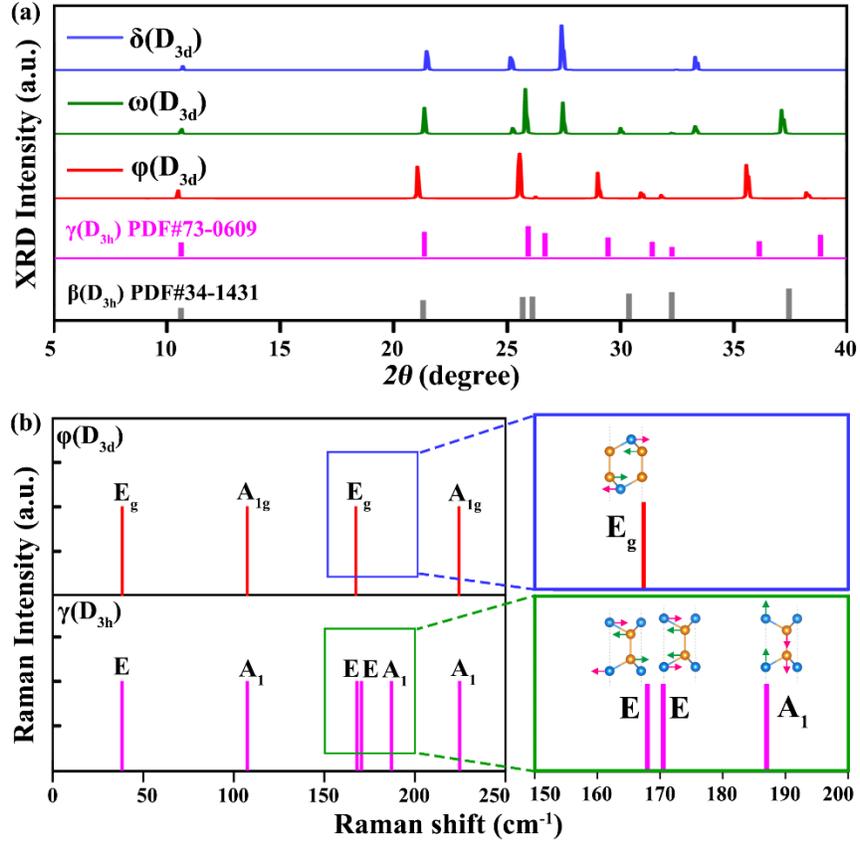

**Figure 5.** (a) XRD patterns of δ($D_{3d}$), ω($D_{3d}$), and φ($D_{3d}$) predicted by theory. The reference PDF data of bulk β($D_{3h}$) and γ($D_{3h}$) are also shown for comparison. (b) Raman spectra of φ($D_{3d}$) and γ($D_{3h}$) phases from DFT simulations (left panel). The zoomed-in part of Raman spectra between 150 and 200 cm$^{-1}$, the inset shows the vibrational modes of the constituent $D_{3d}/D_{3h}$ monolayer (right panel).

# Supporting Information for 'New Polymorphs of Two-Dimensional Indium Selenide with Enhanced Electronic Properties'

*Yuanhui Sun,[†] Yawen Li,[†] Tianshu Li,[†] Koushik Biswas,[‡] Amalia Patanè,[§]*

*and Lijun Zhang[\*,†]*

[†]State Key Laboratory of Integrated Optoelectronics, Key Laboratory of Automobile Materials of MOE and College of Materials Science and Engineering, Jilin University, Changchun 130012, China

[‡]Department of Chemistry and Physics, Arkansas State University, AR 72467, United States

[§]School of Physics and Astronomy, The University of Nottingham, Nottingham NG7 2RD, United Kingdom

*Address correspondence to: lijun_zhang@jlu.edu.cn

# Table of Contents

## 1. Computational methodologies in detail

**Global structure searches.** Searching for the polymorphs of 2D InSe was carried out based on a global minimum search of the free energy landscape with respect to structural variations by combining particle swarm optimization (PSO) algorithm with first-principles energetic calculations.[1,2] The structures of stoichiometric InSe were searched with simulation cell size of 1 – 3 formula units. In the first step, a population of random structures with certain crystallographic symmetry are constructed (the first generation), in which the internal atomic positions are generated by symmetry operations of randomly selected space groups. Then the structures are optimized to the free energy local minima with density functional theory (DFT) calculations. By evaluating the total energy of these structures, 60% of them with lowest enthalpies, together with 40% newly generated structures, are used to produce the structures of next generation by the structure operators of PSO. In this step, a structure fingerprinting technique of band characterization matrix is applied to generate new structures, so that identical (or very similar) structures are strictly forbidden. It significantly enhances the diversity of the generated structures, which is crucial for final convergence of the global structure search. Local structure optimizations are performed by the conjugate gradient method. All of the structure searches reached convergence (*i.e.*, no new structure with lower energy emerging) after 30 generations, at which point ~3600 structures were evaluated.

**First-principles calculations.** The underlying first-principles DFT calculations were carried out by using the plane-wave pseudopotential method as implemented in Vienna *ab initio* Simulation Package.[3,4] The electron-ion interactions were described by the projector augmented wave pseudopotentials[5,6] with In-$5s^25p^1$ and Se-$4s^24p^4$ states treated as valence electrons. We used the generalized gradient approximation formulated by Perdew, Burke, and Ernzerhof[7] as exchange-correlation functional. A kinetic energy cutoff of 520 eV was adopted for wave-function expansion and the *k*-point meshes with spacing $2\pi \times 0.03$ Å$^{-1}$ or less for electronic Brillouin zone integration. A vacuum layer with more than 15 Å thickness was used in layer-dependent calculations to isolate the InSe layer from its neighboring image. The structures (including lattice parameters and atomic positions) were fully optimized after including the essential van der Waals (vdW) interaction until the residual forces were converged within 0.02 eV/Å. The optB86b-vdW functional[8,9] was adopted, that is known to provide a good description of the structural properties β, γ InSe phases.[10] We employed the hybrid functional approach[11] (with 25% exact Fock exchange) to remedy the band gap underestimation in DFT based calculations. The climbing image nudged elastic band method[12,13] was employed to investigate the transition barrier between $D_{3d}$ and $D_{3h}$ monolayers. Five images were used to calculate the reaction path.

*Ab initio* **molecular dynamics and phonon dispersions.** *Ab initio* molecular dynamics simulations were performed at 300 K and 500 K using NVT ensemble and the temperature is controlled by using the Nosé-Hoover method.[14] The total simulation time is 8 ps with a timestep of 2 fs. The phonon calculation of $D_{3d}$ monolayer was performed using a 5×5×1 supercell with the finite displacement approach[15] implemented in Phonopy code.[16] The calculations of δ($D_{3d}$), ω($D_{3d}$), and φ($D_{3d}$) phases were performed using the 5×5×2, 5×5×1, 5×5×1 supercells, respectively.

**Carrier mobility.** On the basis of effective mass approximation, the carrier mobility in 2D materials was calculated within the deformation potential approximation[17] and is expressed as:[10,18–21]

$$\mu_{2D} = \frac{e\hbar^3 C_{2D}}{k_B T m^* m_d (E_l^i)^2}$$

Here, $e$ is the electron charge, $\hbar$ is the reduced Planck constant, $k_B$ is the Boltzmann constant, and $T$ is the temperature (set to 300 K in this paper). $E_l^i$ represents the deformation potential constant of the valence band maximum (VBM) for hole or conduction band minimum (CBM) for electron, defined as $E_l^i = \Delta E_i/(\Delta l/l_0)$. $\Delta E_i$ is energy shift of the $i^{\text{th}}$ band under proper cell dilatation and compression, $l_0$ is equilibrium lattice constant in the transport direction, and $\Delta l$ is the deformation of $l_0$. $m^*$ is effective mass in the transport direction and $m_d$ is the equivalent effective mass determined as $m_d = \sqrt{m_x^* m_y^*}$. The elastic modulus $C_{2D}$ is evaluated by applying longitudinal strain along x and y directions and estimated as $(E - E_0)/S_0 = C_{2D}(\Delta l/l_0)^2/2$, where $E$ and $E_0$ are the total energy under strain and at equilibrium, respectively. $S_0$ is lattice area at equilibrium for the 2D system.

## 2. Structural parameters of predicted new $D_{3d}$ monolayer based polymorphs

**Table S1.** Calculated structural parameters of $\delta(D_{3d})$, $\omega(D_{3d})$, and $\varphi(D_{3d})$.

| Compound | Space group | Lattice parameters (Å, °) | Atomic coordinates (fractional) | | | |
|---|---|---|---|---|---|---|
| | | | Atoms | x | y | z |
| $\delta(D_{3d})$ | P$\bar{3}$m1 | a = b = 4.0838 | In | 0.0000 | 0.0000 | 0.3301 |
| | | c = 8.2409 | Se | 0.3333 | 0.6667 | 0.8262 |
| | | α = β = 90.00 | | | | |
| | | γ = 120.00 | | | | |
| $\omega(D_{3d})$ | P6$_3$mc | a = b = 4.0712 | In | 0.6667 | 0.3333 | 0.1762 |
| | | c = 16.6407 | In | 0.6667 | 0.3333 | 0.3443 |
| | | α = β = 90.00 | Se | 0.0000 | 0.0000 | 0.4223 |
| | | γ = 120.00 | Se | 0.3333 | 0.6667 | 0.0983 |
| $\varphi(D_{3d})$ | R$\bar{3}$m | a = b = 4.0646 | In | 0.0000 | 0.0000 | 0.4447 |
| | | c = 25.3085 | Se | 0.0000 | 0.0000 | 0.2734 |
| | | α = β = 90.00 | | | | |
| | | γ = 120.00 | | | | |

## 3. Energetics of the polymorphs calculated with different vdW functionals

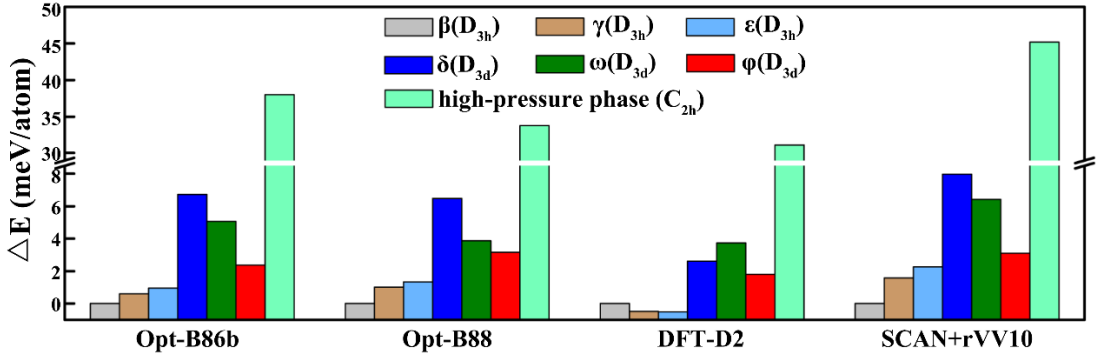

**Figure S1.** Energy differences of InSe-based polymorphs using different vdW functionals.

The energy differences is expressed as: $\Delta E = (E - E_\beta)/N$, in which $E$ is the energy of InSe-based polymorphs, $E_\beta$ is the energy of $\beta(D_{3h})$, and $N$ is the total number of atoms. As shown in Figure S1, calculations using different vdW functionals (opt-B86b,[8,9] opt-B88,[8,9] DFT-D2,[22] and SCAN+rVV10[23]) indicate they have comparable energies within 5 meV/atom. The energy sequence is always $\delta(D_{3d}) > \omega(D_{3d}) > \varphi(D_{3d})$, except for the sequence $\omega(D_{3d}) > \delta(D_{3d}) > \varphi(D_{3d})$ when DFT-D2 functional is used.

## 4. Electronic structures of the polymorphs with varying layer thickness

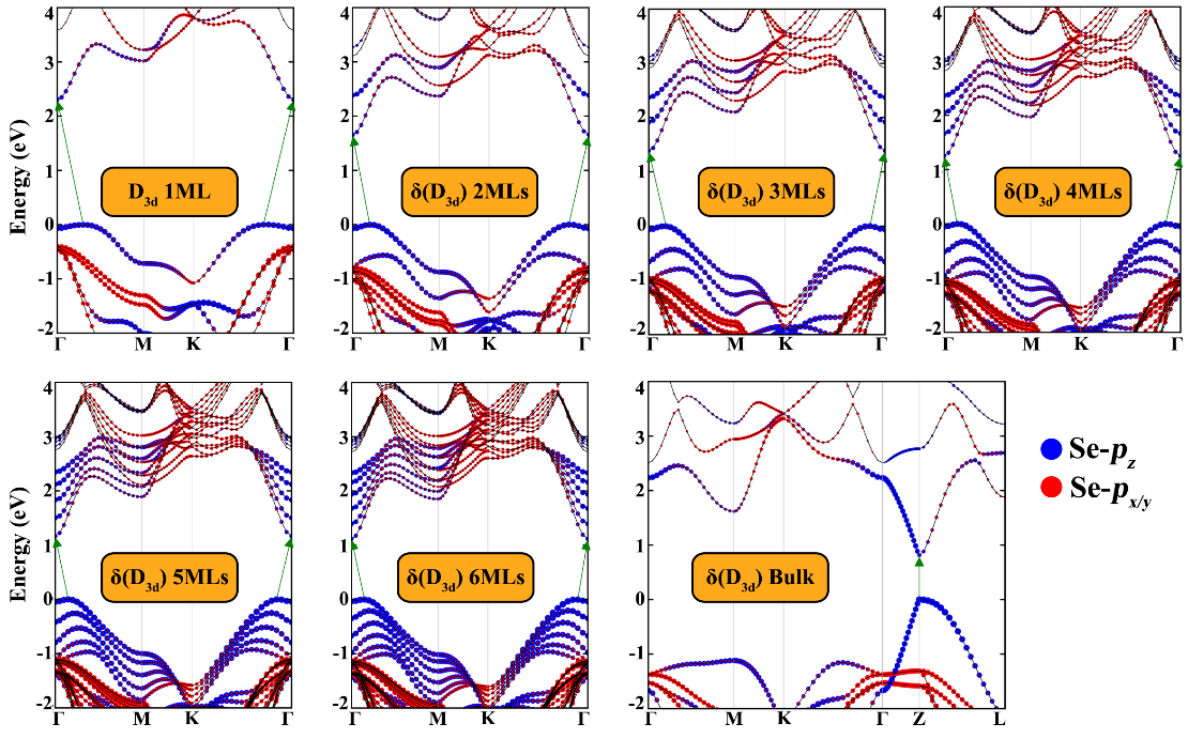

**Figure S2.** The DFT band structures of δ($D_{3d}$) $n$MLs ($n$ = 1-6) and the bulk phase. The band gaps are corrected with scissor operator. The arrows depict the indirect band gap transition for δ($D_{3d}$) $n$MLs and direct band gap transition of bulk phase.

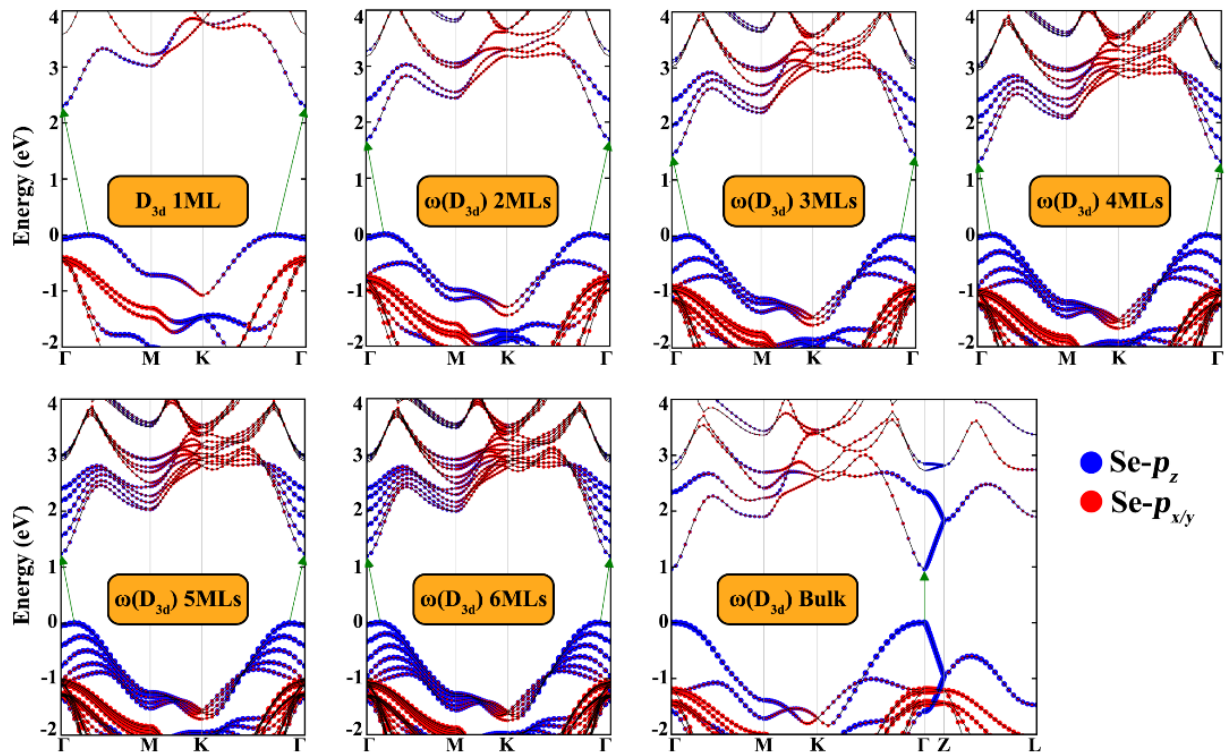

**Figure S3.** The DFT band structures of ω($D_{3d}$) nMLs (n = 1-6) and the bulk phase where the band gaps are corrected with scissor operator. The arrows show the indirect band gap transition for ω($D_{3d}$) nMLs and direct band gap transition of bulk phase.

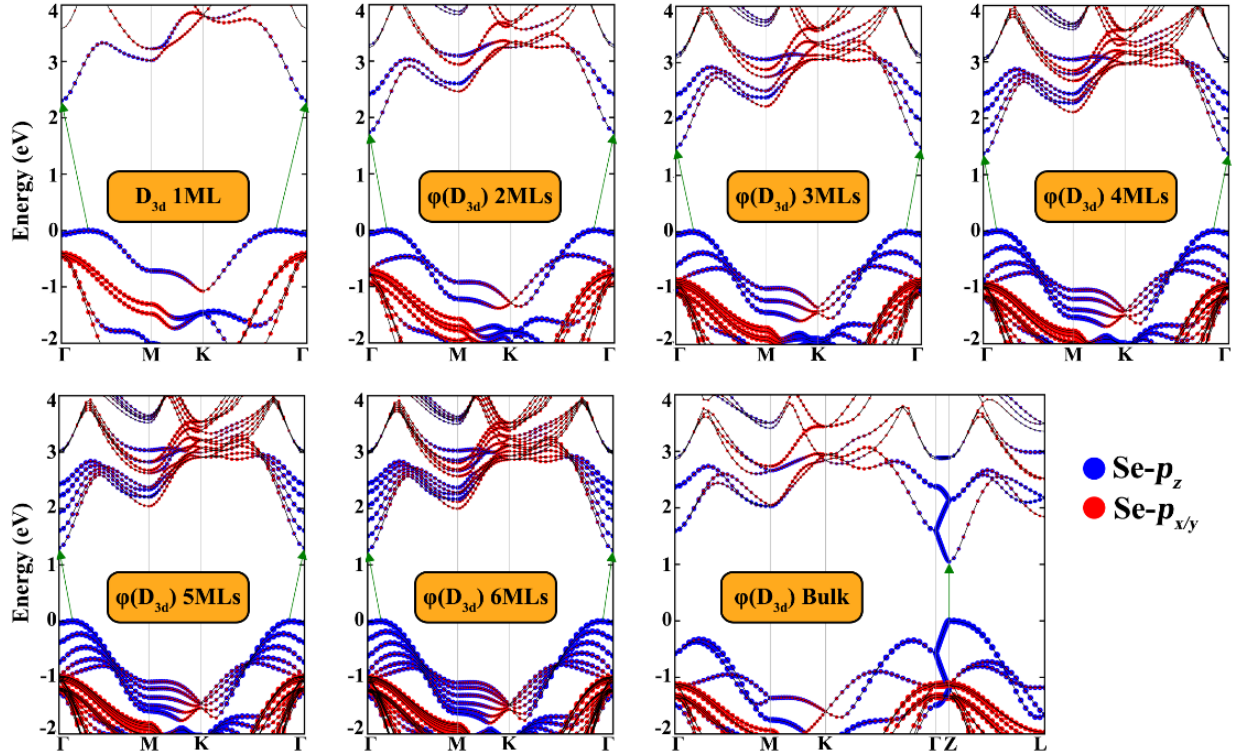

**Figure S4.** The DFT band structures of $\varphi(D_{3d})$ $n$MLs ($n$ = 1-6) and the bulk phase with band gaps corrected using scissor operator. The arrows show the indirect band gap transition for $\varphi(D_{3d})$ $n$MLs and direct band gap transition of bulk phase.

The scissor operator was used to rigidly shift the conduction bands to remedy the band gap underestimation issue of the semilocal DFT calculation. The shift value is the band gap difference between DFT and hybrid functional calculations for the $D_{3d}$ monolayer (ML). With increasing thickness from monolayer to bulk, band edge shift creates direct gap behavior accompanied by considerable decrease in the gap value.

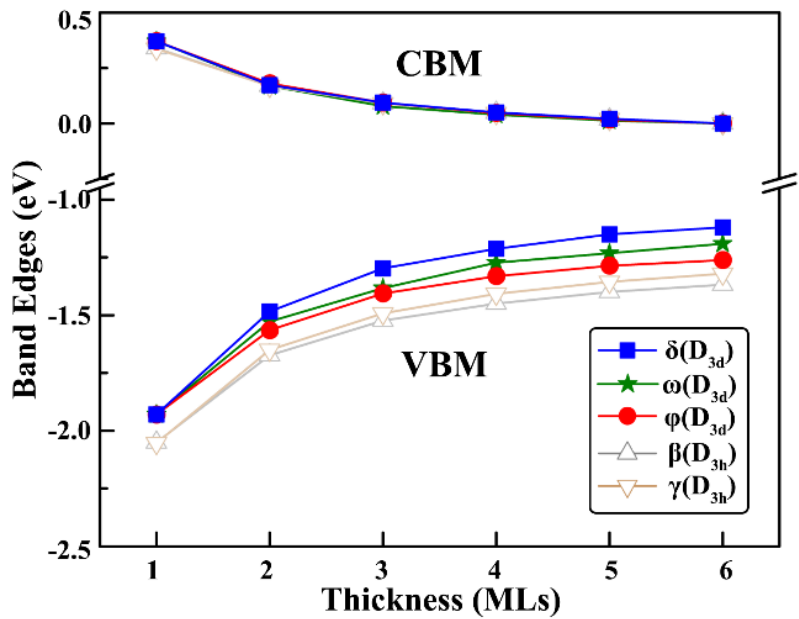

**Figure S5.** Evolution of band edges with increasing number of layers for the three predicted $D_{3d}$ monolayer based polymorphs and two known $D_{3h}$ monolayer based polymorphs.

As shown in Figure S5, the energies of CBMs and VBMs are aligned to the vacuum level and the relevant CBMs of 6MLs structures are set to zero. From the variation tendencies of CBMs and VBMs, it is evident that both band edges contribute to the band gap diminution with increasing layer thickness. However, the variations (upward shift) of the VBMs are much larger than those of CBMs. Among all polymorphs, the $\delta(D_{3d})$ phase shows the largest change in the VBM energy, indicating its strongest interlayer interaction/coupling.

# 5. Carrier mobilities of $D_{3d}/D_{3h}$ monolayer based polymorphs with varying layer thickness

**Table S2.** The electron effective mass ($m_{arm}$, $m_{zig}$), electron deformational potential (eV), 2D elastic modului (N/m) and the corresponding electron mobility (cm$^2$V$^{-1}$s$^{-1}$) along armchair and zigzag directions from monolayer to 6MLs for $D_{3d}$ and $D_{3h}$ monolayer based polymorphs of InSe.

| | $m_{arm}$ | $m_{zig}$ | $E_{arm}$ | $E_{zig}$ | $C_{arm\_2D}$ | $C_{zig\_2D}$ | $\mu_{arm\_2D}$ | $\mu_{zig\_2D}$ |
|---|---|---|---|---|---|---|---|---|
| $D_{3d}$ monolayer | 0.18 | 0.19 | 5.79 | 5.77 | 51.81 | 51.11 | 944.96 | 911.62 |
| 2MLs δ($D_{3d}$) | 0.15 | 0.16 | 5.67 | 5.56 | 100.06 | 110.15 | 2835.94 | 3089.24 |
| 3MLs δ($D_{3d}$) | 0.14 | 0.15 | 5.38 | 5.37 | 160.37 | 159.34 | 5804.67 | 5552.93 |
| 4MLs δ($D_{3d}$) | 0.13 | 0.14 | 5.32 | 5.38 | 208.3 | 215.5 | 8443.80 | 8113.91 |
| 5MLs δ($D_{3d}$) | 0.13 | 0.14 | 5.36 | 5.38 | 271.65 | 267.45 | 11444.61 | 10609.05 |
| 6MLs δ($D_{3d}$) | 0.13 | 0.14 | 5.30 | 5.33 | 317.77 | 309.91 | 14016.06 | 12908.81 |
| 2MLs ω($D_{3d}$) | 0.16 | 0.16 | 5.45 | 5.51 | 99.15 | 97.86 | 2712.65 | 2619.37 |
| 3MLs ω($D_{3d}$) | 0.15 | 0.14 | 5.40 | 5.31 | 141.90 | 148.53 | 4724.05 | 5364.14 |
| 4MLs ω($D_{3d}$) | 0.14 | 0.14 | 5.56 | 5.47 | 192.54 | 199.01 | 6591.48 | 7453.07 |
| 5MLs ω($D_{3d}$) | 0.14 | 0.14 | 5.39 | 5.37 | 242.43 | 237.74 | 9148.25 | 9439.92 |
| 6MLs ω($D_{3d}$) | 0.14 | 0.13 | 5.72 | 6.04 | 310.09 | 292.86 | 10893.47 | 9719.96 |
| 2MLs φ($D_{3d}$) | 0.16 | 0.17 | 5.61 | 5.63 | 99.83 | 99.84 | 2628.46 | 2512.06 |
| 3MLs φ($D_{3d}$) | 0.15 | 0.15 | 5.28 | 5.27 | 149.60 | 149.42 | 5142.79 | 4941.83 |
| 4MLs φ($D_{3d}$) | 0.14 | 0.15 | 5.28 | 5.28 | 201.38 | 198.69 | 7649.27 | 7168.46 |
| 5MLs φ($D_{3d}$) | 0.14 | 0.14 | 5.24 | 5.24 | 243.70 | 244.04 | 9987.77 | 9498.48 |
| 6MLs φ($D_{3d}$) | 0.13 | 0.14 | 5.35 | 5.30 | 300.07 | 301.16 | 12218.66 | 11924.70 |
| $D_{3h}$ monolayer | 0.20 | 0.23 | 5.70 | 5.76 | 52.35 | 52.89 | 801.09 | 689.20 |
| β($D_{3h}$) 2MLs | 0.17 | 0.20 | 5.37 | 5.78 | 100.57 | 100.18 | 2372.77 | 1734.13 |
| β($D_{3h}$) 3MLs | 0.16 | 0.19 | 5.22 | 5.36 | 153.76 | 153.72 | 4313.90 | 3444.57 |
| β($D_{3h}$) 4MLs | 0.15 | 0.18 | 5.67 | 5.95 | 201.33 | 204.14 | 5418.69 | 4157.81 |
| β($D_{3h}$) 5MLs | 0.15 | 0.18 | 5.34 | 5.43 | 254.96 | 255.43 | 7736.44 | 6246.58 |
| β($D_{3h}$) 6MLs | 0.15 | 0.17 | 5.45 | 5.60 | 308.76 | 308.43 | 9255.32 | 7726.56 |
| γ($D_{3h}$) 2MLs | 0.17 | 0.19 | 5.34 | 5.37 | 100.56 | 100.08 | 2461.60 | 2167.55 |
| γ($D_{3h}$) 3MLs | 0.16 | 0.18 | 4.93 | 4.94 | 151.29 | 155.50 | 4889.05 | 4448.69 |
| γ($D_{3h}$) 4MLs | 0.15 | 0.18 | 5.10 | 5.15 | 205.81 | 202.75 | 6846.64 | 5512.10 |
| γ($D_{3h}$) 5MLs | 0.15 | 0.17 | 5.30 | 5.44 | 255.64 | 255.36 | 8102.90 | 6778.92 |
| γ($D_{3h}$) 6MLs | 0.14 | 0.17 | 5.44 | 5.48 | 311.08 | 309.89 | 10379.64 | 8391.39 |

## 6. SHG response of different polymorphs of InSe

**Table S3.** SHG response of newly identified and existing polymorphs of InSe for monolayer and multilayer structures (odd or even layer thickness > 1). "√" and "×" represent SHG-active and SHG-inactive, respectively.

|  | $\delta(D_{3d})$ | $\omega(D_{3d})$ | $\varphi(D_{3d})$ | $\beta(D_{3h})$ | $\varepsilon(D_{3h})$ | $\gamma(D_{3h})$ |
|---|---|---|---|---|---|---|
| **Monolayer** | × | × | × | √ | √ | √ |
| **Even** | × | √ | × | × | √ | √ |
| **Odd** | × | √ | × | √ | √ | √ |

# References

[1]   Y. Wang, J. Lv, L. Zhu, Y. Ma, *Phys. Rev. B* **2010**, *82*, 094116.
[2]   Y. Wang, J. Lv, L. Zhu, Y. Ma, *Computer Physics Communications* **2012**, *183*, 2063.
[3]   G. Kresse, J. Furthmüller, *Comput. Mater. Sci.* **1996**, *6*, 15.
[4]   G. Kresse, J. Furthmüller, *Phys. Rev. B* **1996**, *54*, 11169.
[5]   G. Kresse, D. Joubert, *Phys. Rev. B* **1999**, *59*, 1758.
[6]   P. E. Blöchl, *Phys. Rev. B* **1994**, *50*, 17953.
[7]   J. P. Perdew, K. Burke, M. Ernzerhof, *Phys. Rev. Lett.* **1996**, *77*, 3865.
[8]   J. Klimeš, D. R. Bowler, A. Michaelides, *Phys. Rev. B* **2011**, *83*, 195131.
[9]   Jiří Klimeš and David R Bowler and Angelos Michaelides, *Journal of Physics: Condensed Matter* **2010**, *22*, 022201.
[10]  Y. Sun, S. Luo, X.-G. Zhao, K. Biswas, S.-L. Li, L. Zhang, *Nanoscale* **2018**, *10*, 7991.
[11]  A. V. Krukau, O. A. Vydrov, A. F. Izmaylov, G. E. Scuseria, *The Journal of Chemical Physics* **2006**, *125*, 224106.
[12]  G. Henkelman, B. P. Uberuaga, H. Jónsson, *The Journal of Chemical Physics* **2000**, *113*, 9901.
[13]  G. Henkelman, H. Jónsson, *J. Chem. Phys.* **2000**, *113*, 9978.
[14]  G. J. Martyna, M. L. Klein, M. Tuckerman, *J. Chem. Phys.* **1992**, *97*, 2635.
[15]  K. Parlinski, Z. Q. Li, Y. Kawazoe, *Phys. Rev. Lett.* **1997**, *78*, 4063.
[16]  A. Togo, F. Oba, I. Tanaka, *Phys. Rev. B* **2008**, *78*, 134106.
[17]  J. Bardeen, W. Shockley, *Phys. Rev.* **1950**, *80*, 72.
[18]  P. J. Price, *Annals of Physics* **1981**, *133*, 217.
[19]  Samantha Bruzzone, Gianluca Fiori, *Appl. Phys. Lett.* **2011**, *99*, 222108.
[20]  G. Fiori, G. Iannaccone, *Proceedings of the IEEE* **2013**, *101*, 1653.
[21]  S. Takagi, A. Toriumi, M. Iwase, H. Tango, *IEEE Transactions on Electron Devices* **1994**, *41*, 2357.
[22]  G. Stefan, *Journal of Computational Chemistry* **2006**, *27*, 1787.
[23]  H. Peng, Z.-H. Yang, J. P. Perdew, J. Sun, *Phys. Rev. X* **2016**, *6*, 041005.